\documentclass[a4paper,12pt]{article}
\usepackage{amsfonts}
\usepackage{amsmath}
\usepackage[italian]{babel}
\usepackage[T1]{fontenc}
\usepackage[utf8]{inputenc}

\input{tcilatex}
\begin{document}

\title{Il fisico del neutrino}
\author{Jacopo De Tullio \\
Centro PRISTEM, Universit\`{a} commerciale L. Bocconi}
\date{Ottobre 2013}
\maketitle

\begin{abstract}
In memory of the famous physicist Bruno Pontecorvo, whereof is just
occurred the centenary anniversary of the birth and the twentieth
anniversary of the death, this paper traces the scientific and human
adventure of the italian scientist, from the pioneering studies of nuclear
physics together with his master Enrico Fermi, to the excellent results in
particle physics, in particular his research and insights about the
neutrino. As much attention is paid to the human experience, his
relationship with politics and the choice to embrace the Soviet socialist
model. A short portrait to commemorate a scientist symbol of the twentieth
century. \newline

\indent In ricordo del celebre fisico Bruno Pontecorvo, di cui \`{e} da poco ricorso
il centenario dalla nascita e il ventennale dalla morte, in questo
contributo si ripercorre l'avventura scientifica e umana dello scienziato
pisano, dagli studi pionieristici di fisica nucleare insieme al maestro
Enrico Fermi, agli eccellenti risultati in fisica delle particelle, in
particolare le sue ricerche e le sue intuizione sul neutrino. Altrettanta
attenzione viene rivolta alla sua esperienza umana, il suo rapporto con la
politica e la scelta di abbracciare il modello socialista sovietico. Un
breve ritratto per ricordare uno scienziato protagonista e simbolo del
Novecento.
\end{abstract}

Negli anni `50 del secolo scorso poco lontano da Mosca, sulle rive del
Volga, sorgeva la giovane citt\`{a} di Dubna. In questa cittadina, voluta
nell'immediato dopoguerra dal fisico nucleare ed ex direttore del progetto
per la costruzione della bomba atomica sovietica Igor Kurciatov, era stato
allora costruito il pi\`{u} potente acceleratore di particelle elementari
del mondo e, bench\'{e} gli esperimenti riguardassero soltanto problemi
fisici fondamentali, tutti i lavori erano coperti dal segreto pi\`{u}
assoluto. Tutti i calcoli e le annotazioni venivano riportate solo su un
apposito registro, che veniva poi custodito in un reparto particolarmente
protetto. Ma fra le annotazioni ne spiccavano alcune che riportavano
l'intestazione in russo mentre il testo era in inglese. Come poteva essere
accaduto in un giornale russo \textit{top secret} e di chi era quello
scritto? L'allora direttore dei laboratori V. P. Dzhelepov in merito a quel
periodo afferm\`{o}: \textit{\textquotedblleft Nell'autunno del 1950 venimmo
a sapere che nel nostro laboratorio avrebbe lavorato un noto fisico italiano
allievo del celebre Fermi\textquotedblright } [1]. Si trattava di Bruno
Pontecorvo, che era da poco arrivato a Mosca dall'estero; allora vi furono
molte leggende su come Pontecorvo fosse arrivato in Unione Sovietica e per
quali motivi avesse deciso di stabilirsi in Russia.

A Dubna Pontecorvo prese dimora in una villetta nella via principale della
cittadina scientifica, sia lui che i familiari non conoscevano il russo,
nacquero cos\`{\i} delle difficolt\`{a} che per\`{o} Pontecorvo super\`{o}
con il suo fascino naturale e il modo di comportarsi. In Russia si usa
chiamare per nome patronimico, i colleghi e amici erano a disagio a chiamare
Pontecorvo soltanto Bruno, cos\`{\i} gli chiesero il nome di suo padre e da
allora assunse ufficialmente il nome Bruno Maksimovich (dal nome del padre
Massimo) Pontekorvo.

Bruno Pontecorvo era nato il 22 agosto 1913 a Marina di Pisa da una famiglia
benestante -- suo padre dirigeva una fabbrica di stoffe -- di origine
ebraica e con una certa tendenza alla genialit\`{a}: il fratello maggiore di
Bruno, Guido Pontecorvo, diventer\`{a} biologo e genetista di fama mondiale,
il fratello minore Gillo uno dei pi\`{u} grandi e ammirati registi
cinematografici e fra i cugini ricordiamo Emilio Sereni ed Eugenio Colorni,
il primo partigiano e politico, il secondo filosofo e antifascista.
Particolarmente stimolanti erano anche le frequentazioni della famiglia,
infatti passarono per casa Pontecorvo due giovani studenti di Fisica di nome
Enrico Fermi\footnote{%
Enrico Fermi, nato a Roma il 29 settembre 1901, dopo aver frequentato il
Liceo a Roma vinse il concorso per l'ammissione alla Scuola Normale di Pisa.
Nel 1922 si laure\`{o} in Fisica e una volta tornato a Roma ottenne una
borsa di studio grazie alla quale si rec\`{o} in Germania e Olanda dove entr%
\`{o} in contatto con i maggiori fisici del suo tempo: Heisenberg, Pauli,
Ehrenfest. Tornato in Italia fu professore incaricato a Firenze e poi a Roma
dove il senatore e scienziato Orso Maria Corbino riusc\`{\i} a far istituire
apposta per Fermi la cattedra di Fisica teorica. A Roma Fermi costitu\`{\i}
un gruppo di giovani studenti presso l'Istituto di Fisica di via Panisperna.
Le scoperte fatte a Roma -- fra cui i bombardamenti degli atomi coi neutroni
-- gli valsero il Nobel per la Fisica nel 1938; dopo la consegna del Premio,
a causa della difficile situazione politica italiana, si trasfer\`{\i} negli
Stati Uniti. A New York ottenne una cattedra alla Columbia University e poi
pass\`{o} alla Chicago University dove realizz\`{o} la prima pila atomica e
il primo caso di reazione a catena controllata. Durante la seconda guerra
mondiale Ferm\`{\i} partecip\`{o} a Los Alamos al \textquotedblleft progetto
Manhattan\textquotedblright\ nel quale un gruppo di scienziati produssero la
prima bomba atomica che fu fatta esplodere per la prima volta il 16 luglio
1945 nel deserto del New Mexico. Nel dopoguerra Fermi torn\`{o} spesso in
Italia: a partire dal 1949 fu a Como per partecipare a una conferenza sui
raggi cosmici e a Roma e Milano per un ciclo di lezione organizzate
dall'Accademia dei Lincei; infine nel 1954 insegn\`{o} alla Scuola Estiva di
Varenna. Lo stesso anno si ammal\`{o} e mor\`{\i} a Chicago il 29 novembre
1954.} e Franco Rasetti\footnote{%
Franco Rasetti, nato a Pozzuolo Umbro il 10 agosto 1901, frequent\`{o} la
Scuola Normale di Pisa, dove conobbe Enrico Fermi, laureandosi in Fisica nel
1922. Lo stesso anno ottenne un dottorato annuale di ricerca in California,
dove si dedic\`{o} a esperimenti sullo spettro Raman delle molecole di azoto
che si riveleranno fondamentali per la comprensione di alcune propriet\`{a}
dei nuclei atomici. Nel 1930 torn\`{o} a Roma dove ottenne la cattedra di
spettroscopia all'Universit\`{a} \textquotedblleft La
Sapienza\textquotedblright , trasferendosi nel celebre istituto di Via
Panisperna e iniziando la sua collaborazione con Enrico Fermi. Con il
complicarsi della situazione politica italiana, nel 1939 Rasetti emigr\`{o}
in Canada, presso la Universit\'{e} Laval di Quebec City, dove vi rimase
fino al 1947, compiendo ricerche sui raggi cosmici e di spettroscopia
nucleare. Contrario al coinvolgimento degli scienziati nelle ricerche
belliche rifiut\`{o} di prendere parte al Progetto Manhattan e cominci\`{o}
a dedicarsi agli studi naturalistici, sua grande passione sin dalla
giovinezza. Nel `47 si trasfer\`{\i} alla Johns Hopkins University di
Baltimora, dove insegn\`{o} fisica, geologia, paleontologia, entomologia e
botanica. Nel 1967 fece ritorno a Roma, dove trascorse dieci anni, per poi
trasferirsi a Waremme, in Belgio, paese natale della moglie. Continu\`{o} a
lungo a viaggiare e a fotografare orchidacee, sua ultima fatica scientifica,
fino a che i primi segni della malattia gli limitarono i movimenti. La sua
scomparsa \`{e} avvenuta il 5 dicembre 2001.}. Quest'ultimi Bruno li
incontrer\`{a} nuovamente nel 1931, in veste di commissari, nel colloquio di
ammissione al terzo anno della Facolt\`{a} di Fisica di Roma dopo che aveva
frequentato e superato il biennio di Ingegneria a Pisa. Fermi e Rasetti in
quegli anni stavano cercando di organizzare un gruppo di lavoro sulla Fisica
nucleare, che ai tempi era agli albori, cos\`{\i}, dopo aver superato la
prova di ammissione, il giovane Bruno -- che aveva solo diciotto anni e per
questo fu soprannominato \textquotedblleft cucciolo\textquotedblright\ --
entr\`{o} a far parte del gruppo dei cosiddetti \textquotedblleft ragazzi di
via Panisperna\textquotedblright\ dal nome della via in cui era ospitato
l'Istituto di Fisica.

Bruno era ancora uno studente di Ingegneria, quando sent\`{\i} parlare per
la prima volta del neutrino. L'esistenza di questa speciale particella fu
ipotizzata nel 1930 dal fisico teorico austriaco Wolfgang Pauli che la associ%
\`{o} al fenomeno del \textit{decadimento }$\mathit{\beta }$. Questo
processo fisico \`{e} un particolare tipo di decadimento radioattivo che
coinvolge le forze nucleari deboli, in cui in cui si verifica l'emissione da
parte di un nucleo di un elettrone negativo o positivo e la sua conseguente
trasformazione in un nuovo elemento. Ma la radioattivit\`{a} $\mathit{\beta }
$ in quegli anni poneva un grave problema: gli elettroni non venivano emessi
con una singola energia ma con uno spettro di energie che variava con
continuit\`{a}. A differenza di quanto accadeva nella gran parte dei
fenomeni di decadimento $\mathit{\alpha }$ e $\mathit{\gamma }$, in cui
l'energia della particella emessa \`{e} determinata dalla differenza di
energia tra il nucleo iniziale e quello finale e quindi \`{e} sempre la
stessa, nel caso del decadimento $\mathit{\beta }$ si registrava la
\textquotedblleft scomparsa\textquotedblright\ di una porzione di energia
dai prodotti finali del processo, in palese violazione con le leggi di
conservazione. Per cercare di fornire una soluzione al problema, Pauli afferm%
\`{o} che nel decadimento $\mathit{\beta }$ non veniva emesso soltanto un
elettrone o un positrone, ma anche una seconda particella di natura neutra%
\footnote{%
La particella doveva essere neutra altrimenti sarebbe stata rilevata tramite
il suo potere ionizzante e non poteva essere un fotone poich\'{e} i dati
sperimentali sembravano escluderlo.}, che sfuggiva ai loro strumenti,
portatrice dell'energia mancante.

Pauli chiam\`{o} questa particella \textquotedblleft
neutrone\textquotedblright\ ma quando nel 1932 il fisico inglese James
Chadwik annunci\`{o} di aver scoperto il \textquotedblleft
neutrone\textquotedblright , Pontecorvo e gli altri ragazzi che
frequentavano l'Istituto di Fisica chiesero a Fermi se la particella
scoperta dell'inglese fosse quella ipotizzata da Pauli\footnote{%
Ai tempi si credeva che il nucleo fosse composto soltanto da elettroni e
protoni.}. Fermi rispose che i neutroni di Chadwick erano troppo grandi e
pesanti, mentre i neutroni previsti dal fisico austriaco dovevano essere
piccoli e leggeri, cos\`{\i} ribattezz\`{o} quest'ultimi \textquotedblleft
neutrini\textquotedblright\ [2]. Ma Fermi e il suo gruppo non si limitarono
a dare un nome nuovo alle particelle, proseguirono gli studi fino ad
arrivare a quello che Bruno Pontecorvo defin\`{\i} il \textit{%
\textquotedblleft debutto nel campo della fisica nucleare
pura\textquotedblright }. Nel 1933 Fermi propose la sua teoria del
decadimento $\mathit{\beta }$ [3], la cui pubblicazione fu respinta dalla
rivista \textit{Nature} poich\'{e} conteneva troppe speculazioni astratte e
considerate lontane dalla realt\`{a} fisica, in cui veniva descritto il
neutrino nel suo giusto ambito fisico\footnote{%
L'osservazione diretta di un neutrino non avvenne prima del 1956 con
l'esperimento dei fisici statunitensi Clyde Cowan e Frederick Reines, in cui
neutrini prodotti da un reattore nucleare vennero fatti interagire con i
protoni dell'acqua contenuta in un serbatoio.} e si dimostrava l'esistenza
dell'\textit{interazione debole}, una nuova forza fondamentale della natura
che si aggiungeva alle due fino ad allora note: la forza di gravit\`{a} e la
forza elettromagnetica.

Il `34 \`{e} anche l'anno in cui i ragazzi di via Panisperna, che
comprendevano Edoardo Amaldi ed Emilio Segr\'{e}\footnote{%
Edoardo Amaldi, nato a Carpaneto Piacentino il 5 settembre 1908, figlio del
matematico Ugo, dal 1937 fu professore di fisica sperimentale a Roma
(cattedra che ricopr\`{\i} per oltre 40 anni) dove partecip\`{o} all'attivit%
\`{a} di ricerca del gruppo diretto da Enrico Fermi sulla fisica nucleare e
delle particelle. Nel dopoguerra promosse la realizzazione dei primi
acceleratori di particelle in Italia (elettrosincrotrone di Frascati) e fu
tra i fondatori dell'INFN (di cui fu presidente dal 1960 al 1965), del CERN
di Ginevra e dell'Agenzia Spaziale Europea. Ader\`{\i} inoltre al \textit{%
Pugwash Conferences on Science and World Affairs}, movimento per lo
smantellamento delle armi nucleari. Amaldi \`{e} morto a Roma il 5 dicembre
1989.
\par
Emilio Segr\`{e}, nato a Tivoli il 1 febbraio 1905 da una famiglia di
origini ebraiche, studi\`{o} fisica all'Universit\`{a} di Roma, dove fu
allievo di Fermi e fu membro del gruppo di ricerca dell'Istituto di Fisica
di via Panisperna. Nominato assistente professore di fisica all'Universit%
\`{a} di Roma nel 1932, qui vi rimase fino al 1936. Nel 1938, durante
l'emanazione delle leggi razziali, si trovava all'Universit\`{a} di
Berkeley, dove rimase per il resto della sua vita. Durante la guerra partecip%
\`{o}, insieme a Fermi e Rossi al progetto Manhattan. Nel 1955, lavorando
con Chamberlain all'acceleratore di particelle Bevatron di Berkeley, scopr%
\`{\i} l'antiprotone nelle interazioni protone-nucleone ad alta energia. Per
questa scoperta gli fu conferito nel 1959 il premio Nobel per la fisica. Nel
1974 fu chiamato a ricoprire la cattedra di fisica nucleare all'Universit%
\`{a} di Roma. Segr\`{e} \`{e} morto a Lafayette (California) il 22 aprile
1989.}, iniziarono il celebre esperimento nel quale bombardarono i nuclei
atomici con i neutroni di Chadwik [4][5], ottenendo durante queste prove,
pur non accorgendosene, la fissione di nuclei atomici\footnote{%
Durante gli esperimenti eseguiti da Fermi e i suoi collaboratori era stato
notato che bombardando alcuni nuclei leggeri con i neutroni lenti si
otteneva la formazione di diversi nuclidi radioattivi. Solo nel 1939 i
fisici tedeschi Otto Hann e Fritz Strassmann, che condussero analoghi
esperimenti con l'uranio, provarono che quando questo veniva bombardato con
i neutroni si otteneva la formazione di due grossi frammenti nucleari che
subivano successivamente una serie di trasformazioni radioattive. A questo
tipo di reazione nucleare -- che consiste nella scissione del nucleo atomico
degli elementi pi\`{u} pesanti in due parti (raramente pi\`{u} di due)
aventi masse che stanno in un rapporto dell'ordine di $\frac{3}{2}$\ -- fu
dato il nome di fissione nucleare.}. Questa tecnica innovativa far\`{a} di
Roma, per almeno i successivi quattro anni, la capitale mondiale della
Fisica nucleare.

La natura di Pontecorvo si distingueva da quella degli altri componenti del
gruppo poich\'{e}, oltre a mostrare grandi doti come fisico sperimentale e
teorico, era evidente in lui il profilo di abile fenomenologo, ossia una
grande capacit\`{a} di approfondire applicazioni e ipotesi di lavoro. Fu
nell'estate successiva che Pontecorvo insieme a Edoardo Amaldi, nel
proseguire gli esperimenti sui nuclei atomici, si accorsero che quando i
neutroni attraversavano un filtro di paraffina risultavano cento volte pi%
\`{u} efficaci del solito nel provocare la radioattivit\`{a} dell'argento.
Fermi e il suo gruppo compresero che i neutroni venivano rallentati dalla
paraffina e dunque avevano maggiore probabilit\`{a} di incontrare i nuclei
dell'elemento irradiato e renderlo attivo. Nascevano cos\`{\i} gli
esperimenti coi cosiddetti \textquotedblleft neutroni
lenti\textquotedblright\ che avevano la propriet\`{a} di rimanere nelle
vicinanze del nucleo per un tempo sufficientemente lungo da aumentare la
loro probabilit\`{a} di essere riassorbiti. A questo punto della sua
carriera Bruno Pontecorvo era poco pi\`{u} che ventenne ma era gi\`{a}
entrato nella storia della Fisica.

Nel 1936, dopo gli eclatanti successi, il ventitreenne Pontecorvo si rec\`{o}
a Parigi con una borsa di studio del Ministero per l'Educazione Nazionale.
Qui, grazie a una raccomandazione di Fermi, collabor\`{o} con Fr\'{e}d\'{e}%
ric e Ir\`{e}ne Joliot-Curie -- rispettivamente genero e figlia di Pierre e
Marie Curie e vincitori nel 1935 del premio Nobel per la scoperta della
radioattivit\`{a} artificiale -- su degli esperimenti riguardanti gli urti
tra neutroni e protoni e le transizioni elettromagnetiche tra isomeri. Tra i
risultati ricordiamo la predizione dell'esistenza di isomeri nucleari
stabili dal punto di vista della radioattivit\`{a} $\beta $, confermata nel
1938 quando ne individu\`{o} il primo esempio nel cadmio eccitato da
neutroni veloci. Dalla collaborazione con Andr\'{e} Lazard riusc\`{\i} a
produrre isomeri beta-stabili mediante irradiamento di nuclei stabili con
uno spettro continuo di raggi X di alta energia [6]; a questo effetto Joliot
diede il nome di fosforescenza nucleare.

A Parigi Bruno Pontecorvo, oltre a farsi notare per le sue doti di
scienziato, si fa notare per il suo fisico sportivo da grande giocatore di
tennis e proprio nella capitale francese incontr\`{o} la giovane svedese
Marianne Nordblom che divenne poco dopo tempo sua moglie e da cui ebbe
presto il primo figlio Gil. Ma Parigi gli fu fatale non solo in amore, qui
nel clima del \textit{Front Populaire} e della guerra di Spagna cominci\`{o}
a interessarsi di politica. Gran parte dei suoi colleghi erano di sinistra,
Ir\`{e}ne Joliot era membro del governo del socialista L\'{e}on Blum, il
marito Fr\'{e}d\'{e}ric Joliot attivo comunista, come anche suo cugino
Emilio Sereni, intellettuale e dirigente del PCI, perseguitato in Italia dal
regime fascista e rifugiatosi per l'appunto in Francia. Grazie a suo cugino,
Bruno stabil\`{\i} rapporti con tutta l'\textit{intelligencija politica emigr%
\'{e}e} e nell'agosto del `39, in presenza di Luigi Longo, ader\`{\i} e si
iscrisse al PCI. Dopo l'entrata in vigore delle leggi razziali del 1938
Pontecorvo, ebreo e comunista, era dovuto restare (raggiunto dal fratello
minore Gillo a cui era particolarmente legato) in Francia. Nel settembre del
`39 scoppi\`{o} la guerra e nel giugno del `40, contemporaneamente
all'invasione di Parigi da parte delle truppe tedesche, l'Italia prese parte
al conflitto mondiale. Per gente come Pontecorvo cominciava ormai a non
esserci pi\`{u} posto nel Vecchio Continente e cos\`{\i} decise di lasciare
la Francia; scapp\`{o} in bicletta da Parigi a Tolosa, da dove, con un
rocambolesco viaggio, raggiunse prima la Spagna, poi Lisbona in treno e qui
si imbarc\`{o} per gli Stati Uniti. Nell'agosto del 1940 era con la famiglia
a Tulsa, nell'Oklahoma, dove lavorava in una compagnia petrolifera, la Well
Surveys, in cui mise a punto una tecnica di introspezione dei pozzi
petroliferi basata sul tracciamento di neutroni -- il cosiddetto carotaggio
neutronico\footnote{%
Consiste in una forte sorgente di neutroni (radio+berillio) e di una camera
di ionizzazione, fatta in modo da essere adeguatamente schermata dai raggi
provenienti direttamente dalla sorgente. Come conseguenza dell'interazione
dei raggi primari provenienti dalla sorgente con le formazioni che la
circondano, l'indicazione fornita dalla camera a ionizzazione varia col
variare delle propriet\`{a} degli strati.} dei pozzi di petrolio -- che
rappresentava la prima applicazione della scoperta delle propriet\`{a} dei
neutroni lenti fatta all'Istituto di via Panisperna.

Negli Stati Uniti, probabilmente a causa delle sue idee comuniste, non fu
coinvolto nel Progetto Manhattan per la costruzione della bomba atomica e cos%
\`{\i} nel `43 si trasfer\`{\i} in Canada dove fu chiamato dai laboratori di
Montreal e poi di Chalk River a partecipare a ricerche teoriche nel campo
dei raggi cosmici, delle particelle elementari ad alta energia, in
particolare dei neutrini e del decadimento del muone. Entr\`{o}a far parte
del progetto di costruzione e messa in opera del reattore nucleare NRX a
uranio naturale e acqua pesante, impianto che all'epoca aveva la maggiore
intensit\`{a} ed elevatissimi flussi di neutroni termici. Il 22 luglio 1947
Pontecorvo \`{e} tra i quattro fisici presenti nella sala di controllo
quando il reattore entra in funzione.

Nel 1946 nell'articolo \textquotedblleft Inverse beta
process\textquotedblright\ [7] perfezion\`{o} un metodo radiochimico, basato
sulla trasmutazione cloro-argon, per cercare di catturare i neutrini.
Sebbene la rilevazione del decadimento beta inverso -- durante il quale un
neutrone bombardato con neutrini (o con elettroni e raggi gamma) si
trasforma in protone o viceversa -- era all'epoca considerata impossibile,
Bethe e Peierls nel 1934 avevano pubblicato un lavoro [8] in cui
dimostravano la bassa capacit\`{a} di interazione dei neutrini con la
materia nell'attraversare la Terra (soltanto uno su 1010) e che dunque 
\textit{\textquotedblleft non esiste in pratica alcun modo per osservare il
neutrino\textquotedblright }, secondo Pontecorvo l'uso di potenti fonti di
neutrini come i reattori nucleari, nel suo caso il NRX, poteva realizzare il
processo sopracitato. Il 4 settembre 1946 a Montreal durante il congresso di
fisica nucleare, Pontecorvo afferm\`{o}:\textit{\ \textquotedblleft Il
reattore canadese NRX, del cui progetto stavo facendo parte, non era ancora
entrato in funzione, eppure mi appariva del tutto evidente che sotto lo
schermo compatto, dove la componente molle dei raggi cosmici risultava
considerevolmente affievolita, si poteva disporre di un flusso di circa 1012
neutrini per centimetro quadrato per secondo\textquotedblright\ }[9].

Pontecorvo propose quindi il seguente metodo: se si irradia una grande
quantit\`{a} di $^{37}Cl$ con i neutrini, deve prodursi un nucleo
radioattivo di $^{37}Ar$ secondo la reazione: $\nu +$ $^{37}Cl$ $\rightarrow 
$\ $^{37}Ar+e$. L'argon 37 decade attraverso un processo detto di cattura
elettronica, secondo cui un elettrone K viene catturato dal nucleo
trasformandolo in un nucleo di carica Z-1 con un periodo di decadimento di
34 giorni. Pontecorvo indicava tra le possibili sorgenti di neutrini i
reattori nucleari (attraverso il decadimento dei prodotti di fissione) il
Sole (attraverso le reazioni di fusione e di decadimento) [10]. Ma il
neutrino \`{e} una particella estremamente elusiva e catturarlo non \`{e}
semplice. La tecnica di Pontecorvo non era perfetta ma partendo dalla
reazione studiata dal fisico italiano, verso la fine degli anni Sessanta
nella miniera di Homestake negli Stati Uniti, per iniziativa di Raymond
Davis (premio Nobel per la Fisica nel 2002), ebbe inizio l'esperimento [11]
che port\`{o} alla cattura dei neutrini solari e per primo a una stima del
loro numero\footnote{%
Il problema dei neutrini solari nacque a partire negli anni `70 dopo la
pubblicazione dei risultati dell'esperimento di Homestake e riguardava una
grossa discrepanza tra il numero osservato di neutrini che arrivano sulla
Terra e il numero predetto da modelli teorici. Il problema \`{e} stato
risolto grazie alle scoperte nell'ambito della Fisica dei neutrini, che
hanno richiesto una modifica del Modello Standard della Fisica delle
Particelle in modo che fossero permesse le oscillazioni dei neutrini.}.

Fra il 1944 e il 1945 Marcello Conversi, Ettore Pancini e Oreste Piccioni
portarono a termine un esperimento [12] di estrema importanza -- che diede
il via alla cosiddetta Fisica delle particelle elementari -- durante il
quale identificarono una particolare particella presente nei raggi cosmici:
il mesotrone (detto oggi mesone $\mathit{\mu }$ o muone). Il muone appariva
come un elettrone duecento volte pi\`{u} pesante, instabile e con una vita
abbastanza corta (prodotto nell'alta atmosfera \textquotedblleft
vive\textquotedblright\ il tempo necessario per arrivare sulla Terra
viaggiando quasi alla velocit\`{a} della luce) che per\`{o} una volta morto
non si disintegrava in un elettrone e un fotone ma solamente in un
elettrone. Questo elettrone inoltre non possedeva un'energia ben definita ma
assortita in un intervallo continuo, il che significava che il muone si
divideva in un elettrone e in almeno altre due particelle neutre invisibili
alle strumentazioni. In questo ambito, nel `47 Pontecorvo fu tra i primi a
dedurre, servendosi delle teorie di Enrico Fermi che avevano introdotto i
neutrini nelle interazioni deboli, che la cattura del muone da parte del
nucleo atomico, proprio come la cattura dell'elettrone, producesse neutrini
[13]. In una lettera a Giancarlo Wick del 1947 [14], scrisse:

\textit{Deep River, 8 maggio 1947}

\textit{Caro Giancarlo }(...)\textit{\ se ne deduce una similarit\`{a} tra
processi beta e processi di assorbimento ed emissione di mesoni, che,
assumendo non si tratti di una coincidenza, sembra di carattere fondamentale.%
}

La simmetria muone-elettrone, messa in evidenza da Pontecorvo, costituiva il
primo indizio dell'esistenza di una universalit\`{a} delle interazioni
deboli scoperte da Fermi, che si rivolgeva dunque a un ambito molto pi\`{u}
generale di quello del decadimento $\mathit{\beta }$. L'idea di universalit%
\`{a} dei decadimenti deboli venne espressa nel 1948 da Giampiero Puppi nel
suo articolo \textquotedblleft Sui mesoni dei raggi
cosmici\textquotedblright\ [15] ed esplicitata graficamente tramite il
cosiddetto \textquotedblleft triangolo di Puppi\textquotedblright . In
questa rappresentazione il decadimento beta, il decadimento del muone e la
cattura nucleare del muone rappresentano i tre lati del triangolo, sono perci%
\`{o} processi accomunati dalla stessa costante di accoppiamento.
Nell'ipotizzare un'uguaglianza tra le costanti di accoppiamento di elettroni
e muoni ai nucleoni si assumeva dunque l'esistenza di una universalit\`{a}
dell'interazione debole. \`{E} possibile affermare che, con la sua
intuizione, Pontecorvo aveva contribuito alla costruzione di due dei lati
del triangolo di Puppi.

Inoltre Pontecorvo, sulla scia delle intuizioni di Ettore Majorana e Gian
Carlo Wick, ipotizz\`{o} che i due neutrini prodotti dal muone fossero di
natura diversa, uno come partner fisso del muone primario, l'altro come
partner dell'elettrone\footnote{%
Sulla pietra tombale di Bruno Pontecorvo nel Cimitero degli Inglesi a Roma
compare la scritta:
\par
$\upsilon _{\mu }\neq \upsilon _{e}$ (\textit{il neutrino-mu \`{e} diverso
dal neutrino-e}).}. A questo punto Bruno Pontecorvo si proponeva come uno
dei pi\`{u} grandi esperti al mondo di Fisica del neutrino.

Nel 1948, dopo aver ottenuto la cittadinanza britannica, Pontecorvo, su
invito del fisico John Cockcroft (che nel 1932, insieme all'irlandese Ernest
Walton, ottenne la disintegrazione dei nuclei di litio e boro mediante
protoni accelerati, esperimento che valse loro il Nobel nel `51) si trasfer%
\`{\i} ad Harwell nei pressi di Oxford. Qui aveva sede l'\textit{Atomic
Energy Research Establishment}, il principale centro di ricerche nucleari
installato dal governo inglese, dove Pontecorvo partecip\`{o} al progetto
per la costruzione della bomba atomica inglese e si dedic\`{o} agli studi
sui raggi cosmici.

Pontecorvo non aveva ancora quarant'anni, era di bell'aspetto, sorridente e
affabile, amante dello sport, ottimo giocatore di tennis (lo era stato
sempre), appassionato di pesca subacquea e di sci nautico. Una condizione
invidiabile se non fosse che il suo mestiere lo collocava nell'ambito
scottante dell'energia atomica per uso bellico e i connessi segreti,
nonostante lui non fu mai coinvolto in maniera diretta. Il fisico italiano,
in occasione delle sue numerose trasferte scientifiche, aveva conosciuto sia
Alan Nunn May sia Klaus Fuchs\footnote{%
Alan Nunn May \`{e} stato un fisico britannico che durante la seconda guerra
mondiale fu una spia sovietica che rivel\`{o} i segreti atomici inglesi, fu
condannato ai lavori forzati nel 1946.
\par
Klaus Fuchs \`{e} stato un fisico teorico tedesco emigrato nel Regno Unito
con l'avvento del nazismo. Fu arrestato nel 1950 dagli agenti di Scotland
Yard con l'accusa di aver ceduto all'URSS alcuni segreti sulla bomba atomica
e all'idrogeno.}, due fisici condannati in Inghilterra per spionaggio in
favore dell'Unione Sovietica; \`{e} in questo oppressivo contesto di
\textquotedblleft caccia alle streghe\textquotedblright\ che si collocano le
scelte e le azioni di Pontecorvo.

Nell'estate del 1950 lo scienziato, insieme a sua moglie e ai suoi tre
figli, lasci\`{o} la casa di Abington, nei pressi di Harwell, senza
avvertire nessuno. Ogni ricerca fu vana e in Parlamento il Ministro per
l'Approvvigionamento militare George Strauss, dopo aver escluso che
Pontecorvo avesse mai avuto accesso a ricerche di carattere segreto, si
disse convinto che si trovasse in Russia [16]. In realt\`{a} lo scienziato
con la sua famiglia aveva raggiunto in macchina l'Italia. Dopo un breve
periodo nella sua terra natale, prese un aereo da Roma con destinazione
Stoccolma e da l\`{\i} si imbarc\`{o} per Helsinki; destinazione successiva
Leningrado. Per superare la cortina di ferro i Pontecorvo si divisero:
Marianne e i ragazzi su un'automobile, Bruno nascosto nel bagagliaio di
un'altra. Entrati in Unione Sovietica e giunti a Mosca gli assegnarono un
comodo appartamento in via Gorkij. I sovietici si mostrarono gentili e
deferenti ma altrettanto inflessibili per quanto riguardava la segretezza.
Per alcuni mesi l'intera famiglia fu costretta all'isolamento e quando il
fisico italiano chiese di poter spiegare alla radio le motivazioni della sua
fuga in URSS glielo vietarono.

Pontecorvo fu trasferito a Dubna, citt\`{a} a un centinaio di chilometri
dalla capitale dove risiede l'aristocrazia della Fisica sovietica, dove gli
fu affidata la direzione della divisione di Fisica sperimentale del
Laboratorio dei Problemi Nucleari. Qui maturarono le sue fondamentali
ricerche nella Fisica delle particelle elementari e, successivamente, in
Astrofisica. Dal `54 al `57 si dedic\`{o} all'interazione di pioni (mesoni pi%
\`{u} leggeri) con i nuclei e, in una pubblicazione del 1959 per primo
dimostr\`{o} per via teorica l'esistenza di diversi tipi di neutrini (come
ipotizzato nel `47) associati ai leptoni carichi (elettroni e muoni) le cui
differenti propriet\`{a} sono rilevabili. Con questo risultato nasceva la
Fisica dei neutrini ad alta energia. Bench\'{e} l'acceleratore di particelle
di Dubna fosse fra i pi\`{u} potenti al mondo (era possibile accelerare i
protoni fino a un'energia di 460 MeV) non era adatto a provare i risultati
di Pontecorvo. Soltanto pochi anni dopo, agli inizi degli anni Sessanta, gli
americani Leon Ledermann, Melvin Schwartz e Jack Steinberger confermarono
sperimentalmente le ipotesi del fisico italiano. Questa scoperta valse ai
tre fisici il premio Nobel per la Fisica nel 1988 per \textit{%
\textquotedblleft il metodo del fascio di neutrini e la dimostrazione della
struttura doppia dei leptoni attraverso la scoperta del neutrino
muone\textquotedblright }, suscitando lo scalpore di una parte della comunit%
\`{a} scientifica internazionale per l'esclusione del fisico teorico
italiano che per primo effettu\`{o} la previsione. In questi anni Pontecorvo
propose anche un metodo per rivelare gli antineutrini prodotti nei reattori
nucleari; questa tecnica fu utilizzata con successo nel 1956 a Savannah
River negli Stati Uniti da Frederick Reines (che per questo esperimento
ricevette nel 1995 il Nobel).

Cittadino sovietico dal 1952, l'anno seguente ricevette il Premio Stalin e
dal `58 fu ammesso all'Accademia sovietica delle scienze. Soltanto nel 1955
torn\`{o} ad apparire in pubblico in occasione di una conferenza stampa
nella sede dell'Accademia, in cui il fisico raccont\`{o} la sua vicenda e le
motivazioni dell'adesione al modello comunista. Il discorso era di carattere
fortemente politico, accus\`{o} duramente gli Stati Uniti di essere una
potenza belligerante rivendicando il ruolo di potenza di pace per l'Unione
Sovietica; in quella stessa occasione ribad\`{\i} la sua estraneit\`{a} a
ogni progetto di costruzione di una centrale atomica. Negli anni a seguire
le sue posizioni politiche rimasero statiche: nel `56 consider\`{o}
\textquotedblleft controrivoluzionari\textquotedblright\ i patrioti di
Budapest e nel `68, pur condannando l'invasione sovietica della
Cecoslovacchia, non ader\`{\i} alle proteste dei comunisti italiani -- che
gli sospesero l'abbonamento al giornale \textit{l'Unit\`{a}} -- e non si
schier\`{o} nella vicenda del fisico dissenziente Sacharov\footnote{%
Andrej Dmitrievi\v{c} Sakharov \`{e} stato un fisico sovietico. Negli anni
`50 contribu\`{\i} alla costruzione della bomba all'idrogeno ma, a partire
dagli anni '70, divenne voce critica nei confronti degli aspetti repressivi
del regime sovietico, tanto da fondare nel 1970 il \textquotedblleft
Comitato per i diritti civili e prendere le difese dei dissidenti e dei
perseguitati\textquotedblright . La sua attivit\`{a} in favore dei diritti
civili fu premiata col Nobel per la Pace nel 1975.}.

Nel 1957 Pontecorvo pubblic\`{o} l'articolo \textquotedblleft Mesonium and
Antimesonium\textquotedblright\ [17] in cui elabor\`{o} la teoria del
mescolamento leptonico. I leptoni sono particelle subatomiche suddivise in
tre famiglie: gli elettroni, i muoni e i tauoni (quest'ultimi sconosciuti a
Pontecorvo perch\'{e} individuati soltanto nella seconda met\`{a} degli anni
`70 da Martin Lewis Perl) a cui \`{e} associato un particolare neutrino. A
ciascun doppietto di leptoni viene assegnato un numero leptonico -- che
tiene conto dei neutrini associati alle particelle -- che viene
rigorosamente conservato (cio\`{e} la somma dei numeri leptonici di ogni
famiglia nello stato iniziale \`{e} uguale alla somma dei numeri leptonici
nello stato finale) in tutte le interazioni. Secondo la teoria elaborata da
Pontecorvo i diversi tipi di neutrini, nel vuoto, possono \textquotedblleft
oscillare\textquotedblright\ ossia trasformarsi gli uni negli altri e quindi
avere una massa, seppure molto piccola. In un articolo del 1969 [18], in
collaborazione con il fisico teorico Vladimir Gribov, presenta in dettaglio
il formalismo matematico della teoria delle oscillazioni. Questo fenomeno --
noto come \textquotedblleft oscillazione dei neutrini\textquotedblright\ --
fu proposto come soluzione ai problemi sorti nella misurazione dei i
neutrini solari del 1968 e fu fecondo di sviluppi. Ma all'epoca la gran
parte della comunit\`{a} scientifica sosteneva che i neutrini fossero
particelle prive di massa. Pontecorvo riprese le sue argomentazioni in vari
articoli in collaborazione con S. M. Bilenky, pubblicati a partire dal 1976:
\textquotedblleft Quark-lepton analogy and neutrino
oscillations\textquotedblright\ [19], \textquotedblleft The lepton-quark
analogy and muonic charge\textquotedblright\ [20], \textquotedblleft
Oscillations in neutrino beams: status and possibilities of
observation\textquotedblright\ [21]\ e \textquotedblleft Again on neutrino
oscillations\textquotedblright\ [22]. In particolare quest'ultimo riassumeva
quali erano le questioni principali a cui dare risposta relativi ai fenomeni
di oscillazione: masse finite dei neutrini, mixing dei neutrini, violazione
della carica leptonica, numero dei tipi di neutrini [23].

Il fenomeno dell'oscillazione dei neutrini \`{e} stato parzialmente
confermato nel `98 con l'esperimento Super-Kamiokande, svolto nella miniera
giapponese di Kamioka, che ha mostrato con una certa attendibilit\`{a} la
presenza di trasformazioni di neutrini atmosferici $\mu $, recentemente \`{e}
stato confermato con gli esperimenti del 2010 condotti dai Laboratori
dell'Istituto Nazionale di Fisica Nucleare del Gran Sasso [24] sui fasci di
neutrini prodotti da acceleratori in cui si \`{e} osservata, in particolare,
l'oscillazione dei neutrini muonici in neutrini $\tau $. Purtroppo
Pontecorvo non ha potuto assistere all'affermazione delle sue straordinarie
intuizioni che sono alla base della fisica del nuovo millennio.

Per molti anni Pontecorvo non pot\'{e} lasciare l'URSS e riusc\`{\i} a
ritornare la prima volta in Italia nel 1978 in occasione del settantesimo
compleanno di Edoardo Amaldi; in quello stesso anno comparvero i primi
sintomi del morbo di Parkinson che progressivamente, senza mai togliergli
lucidit\`{a}, gli intralcer\`{a} i movimenti. Dopo qualche anno torn\`{o} a
stabilirsi a Roma a casa della sorella Laura; sono anni difficili per il
Pontecorvo \textquotedblleft comunista\textquotedblright : una perdita di
fiducia nei dogmi e negli ideali di una vita intera, la disillusione e il
rammarico per la fine dell'Unione Sovietica. In un libro-intervista con
Miriam Mafai [25]\ alla domanda della giornalista se si fosse pentito della
scelta fatta quarant'anni prima, Pontecorvo rispose: \textit{%
\textquotedblleft Ci ho pensato molto, a questa domanda. Puoi immaginare
quanto ci ho pensato. Ma non riesco a dare una risposta}\textquotedblright .

L'amore per la sua seconda patria non lo abbandon\`{o} e cos\`{\i} nel `93
volle tornare a Dubna. Qui, a causa del Parkinson, sub\`{\i} una brusca
caduta dalla bicicletta e si ruppe il femore; decise di curarsi a Roma ma il
suo fisico non resse lo stress e mor\`{\i} a Dubna il 24 settembre 1993.
Bruno Pontecorvo \`{e} stato il massimo scienziato nel campo della Fisica
dei neutrini e tutte le teorie e ipotesi da lui esposte si sono rivelate
corrette; ma \`{e} stato anche un protagonista del Novecento, degli anni
della guerra fredda, dominati da passioni ingenue e da certezze sfrontate.
Anche dopo la sua morte il fisico non ebbe una sola patria e per sua volont%
\`{a} met\`{a} delle ceneri furono sepolte nel cimitero di Dubna e l'altra
met\`{a} riposano nel cimitero acattolico di Roma.

\end{document}